\documentclass[floatfix,aps,twocolumn,showpacs,preprintnumbers,amsmath,amssymb]{revtex4-1}

\usepackage[applemac]{inputenc}
\usepackage[T1]{fontenc}
\usepackage{amsfonts, amssymb, amsmath, dsfont}
\usepackage{graphicx}
\usepackage{lipsum}
\usepackage{float}
\usepackage{color}
\usepackage{ulem}

\newcommand{\bcalM}{\boldsymbol{\mathcal{M}}}
\newcommand{\bcalP}{\boldsymbol{\mathcal{P}}}
\newcommand{\Id}{\boldsymbol{\mathrm{Id}}}
\newcommand{\bF}{\mathbf{F}}
\newcommand{\PI}{\mathbf{\Pi}}
\newcommand{\bLamb}{\mathbf{\Lambda}}
\newcommand{\bPhi}{\mathbf{\Phi}}

\newcommand{\fo}{\mathbf{f}_0}
\newcommand{\go}{\mathbf{g}_0}

\newcommand{\bE}{\mathbf{E}}

\newcommand{\bB}{\mathbf{B}}

\newcommand{\bD}{\mathbf{D}}
\newcommand{\bP}{\mathbf{P}}
\newcommand{\bM}{\mathbf{M}}
\newcommand{\bG}{\mathbf{G}}

\newcommand{\bH}{\mathbf{H}}

\newcommand{\bN}{\mathbf{N}}

\newcommand{\br}{\mathbf{r}}
\newcommand{\bx}{\mathbf{x}}
\newcommand{\by}{\mathbf{y}}
\newcommand{\bz}{\mathbf{z}}

\newcommand{\bp}{\mathbf{p}}
\newcommand{\bbm}{\mathbf{m}}

\def\d{\mathrm{d}}
\def\eps{\varepsilon}
\def\Re{\mathrm{Re}}
\def\Im{\mathrm{Im}}

\def\tot{\mathrm{tot}}

\begin{document}

\title{A chiral route to pulling optical forces and left-handed optical torques}

\author{Antoine Canaguier-Durand}
\email[Electronic address: ]{antoine.canaguier@espci.fr}
\altaffiliation{Present address: Institut Langevin (ESPCI, CNRS), 1 rue Jussieu, 75005 Paris, France.}
\affiliation{ISIS \& icFRC, University of Strasbourg and CNRS, 8 all\'{e}e Gaspard Monge, 67000 Strasbourg, France.}
\author{Cyriaque Genet}
\email[Electronic address: ]{genet@unistra.fr}
\affiliation{ISIS \& icFRC, University of Strasbourg and CNRS, 8 all\'{e}e Gaspard Monge, 67000 Strasbourg, France.}

\begin{abstract}
We analyze how chirality can generate pulling optical forces and left-handed torques by cross-coupling linear-to-angular momenta between the light field and the chiral object. In the dipolar regime, we reveal that such effects can emerge from a competition between non-chiral and chiral contributions to dissipative optical forces and torques, a competition balanced by the strength of chirality of the object. We extend the analysis to large chiral spheres where the interplay between chirality and multipolar resonances can give rise to a break of symmetry that flips the signs of both optical forces and torques. 
\end{abstract}

\maketitle

\section*{Introduction}

Recent work has revealed how specifically tailored light fields can lead to surprising effects in optical transport, such as the definition of long-range tractor beams and thereby optical forces that can ``pull'' illuminated particles towards the source of light \cite{saenz2011optical,chen2011optical,sukhov2010concept,sukhov2011negative,novitsky2011single,novitsky2012material,dogariu2013optically,shvedov2014long,novitsky2014pulling}. Such anomalous pulling forces stem from a strong forward scattering induced when a non-paraxial incident field excites  multipolar moments of a particle. These moments then interfere to generate a recoil term driving a ``backward'' effect that acts against the dipolar contribution to the force. At the lowest order where it writes as a product of electric and magnetic dipoles, the recoil term is smaller than the dipolar contribution by a factor $(r / \lambda)^3$ where $\lambda$ is the illumination wavelength and $r$ the size of the particle. In this framework therefore, pulling forces could only be induced for dielectric particles not too small. 

Chiral light-matter interactions have recently caught attention from which new types of optical forces have been described \cite{canaguier2013mechanical,cameron2014discriminatory,cameron2014diffraction,tkachenko2014optofluidic,canaguier2014chiral} with potential applications in optical enantioseparation. In this context, it has lately been recognized that the mixed electric-magnetic polarizability intrinsic to a chiral object can contribute to the recoil effect \cite{wang2014lateral}. Within circularly polarized standing waves, transferring angular momentum of light into linear momentum of the chiral particle can even reverse the direction of motion of extended specific chiral structures depending on their extension \cite{ding2014realization}. 

In this article, we demonstrate that a backward scattering force can be obtained in a strictly dipolar regime under single plane wave illumination considering a chiral Rayleigh particle. We hence reveal the universal character of pulling effects in the context of chirality by releasing the two requirements of large objects and non-paraxiality. Using the Lorentz law for evaluating the optical force on a chiral dipole, we provide a unified framework which describes, in a most direct way, the linear and angular momentum transfers mediated by the chiral polarizability. One important outcome of our work is to show that these linear-to-angular crossed momentum transfers can take both the form of optical pulling forces and of so-called left-handed optical torques. 

An other appealing aspect of our dipolar approach is the possibility to evaluate explicitly these momentum transfers. We do so by exploiting a quasi-static model for the chiral dipole. Pulling forces and left-handed torques emerge from the competition between non-chiral and chiral contributions to dissipative optical forces as perfectly revealed with a circularly polarized plane wave. A quasi-static modeling of the chiral response shows in a straightforward way how this competition is balanced by the strength of the chiral response of the system to the electromagnetic field. From this approach, chirality appears as a clear alternative to existing proposal in the context of pulling and left-handed dynamic, just like chirality is opening a new window on negative refraction \cite{pendry2004chiral,zhang2009negative}. 

These effects are obviously modified beyond the dipolar regime. For larger objects, higher-order multipolar moments of the optical response of the particle will start playing a role. Although crucial, the consequences of such high order terms has not been discussed yet. To do so, we develop in this article a multipolar approach for evaluating optical forces and torques exerted on large spherical objects. Our approach is based on a generalization of the Mie theory to chiral scatters in order to account for their chiral scattering properties. We show that the crossed momentum transfers discussed in the dipolar regime are indeed strongly modified due to multipolar effects. The interplay between multipoles progressively excited as the radius of the chiral sphere increases and the chiral optical forces and torques lead to a surprising break of symmetry that gives rise to flipping the signs of the optical forces and torques.

%---------------------
\section{Light-matter momentum transfers for a chiral dipole in a chiral field}
%---------------------

\subsection{Linear and angular momentum densities for a chiral harmonic field}

In the absence of charge and current, the energy of a harmonic electromagnetic field $(\bE_0(\br),\bH_0(\br))$ is characterized by a time-averaged density $W(\br)$ and flow $\PI(\br)$
\begin{align}\label{def_energy}
W(\br) &= \frac{\eps_m}{4} \| \bE_0 \|^2 + \frac{\mu_m}{4} \| \bH_0 \|^2 = W_E + W_H \nonumber \\
\PI(\br) &= \frac{1}{2} \Re \left[ \bE_0 \times \bH_0^* \right] = \PI_O + \PI_S.
\end{align}
where $\eps_m$ is the (real) electric permittivity and $\mu_m $ is the (real) magnetic permeability of the medium. Its real refractive index is $n_m = \sqrt{\eps_m \mu_m} / \sqrt{\eps_0 \mu_0}$, where  $\eps_0$ is the vacuum electric permittivity and $\mu_0$ the vacuum magnetic permeability. The energy density is split in electric and magnetic parts, the Poynting vector $\PI(\br)$ is separated into orbital and spin components. These components can be written either in terms of electric or magnetic fields 
\cite{berry2009optical,bekshaev2013subwavelength,bliokh2013dual,canaguier2013force}:
\begin{align} \label{PI_decompo}
\PI &= ~ \underbrace{\frac{\Im[\fo^*]}{2\omega \mu_m}}_{\PI_O^{(E)}} ~ + ~  \underbrace{\frac{\nabla \times \bPhi_E}{2 \omega \mu_m}}_{\PI_S^{(E)}}~ 
= ~ \underbrace{\frac{\Im[\go^*]}{2\omega \eps_m}}_{\PI_O^{(H)}} ~ +  ~  \underbrace{\frac{\nabla \times \bPhi_H}{2 \omega \eps_m}}_{\PI_S^{(H)}},
\end{align}
where we have introduced the vector fields $\fo(\br)= \left( \bE_0 \cdot \nabla \right) \bE_0^* + \bE_0 \times \left( \nabla \times \bE_0^* \right)$ and $\go(\br)=\left( \bH_0 \cdot \nabla \right) \bH_0^* + \bH_0 \times \left( \nabla \times \bH_0^* \right)$, as well as the electric and magnetic ellipticities $\bPhi_E(\br) = - \Im[\bE_0 \times \bE_0^*]/2$ and $\bPhi_H(\br) = - \Im[\bH_0 \times \bH_0^*]/2$. From the $(\bE \leftrightarrow \bH)$ symmetry, this separation can also be given in a dual-symmetric way as:
\begin{align}
\PI_O &= \frac{\PI_O^{(E)}+\PI_O^{(H)}}{2} = \frac{\omega}{4 k^2} \Im\left[ \eps_m \fo^* + \mu_m \go^*\right]  \label{PI_O_dual} \\
\PI_S &= \frac{\PI_S^{(E)}+\PI_S^{(H)}}{2} = \frac{\omega}{4 k^2} \nabla \times  \left[ \eps_m \bPhi_E +\mu_m \bPhi_H \right]  \label{PI_S_dual} 
\end{align}
with $k=\omega n_m / c$.

The chirality of a harmonic field is a conserved quantity, also characterized by a density and a flow. These quantities are time-independent and write as:
\begin{align}\label{def_energy}
K(\br) &= \frac{k^2}{2 \omega} \Im \left[ \bE_0 \cdot \bH_0^*\right] \nonumber \\
\bPhi(\br) &= \frac{\omega}{2} \left[ \eps_m \bPhi_E +\mu_m \bPhi_H \right] 
\end{align}
where the latter quantity appeared in the dual-symmetry spin part of the Poynting vector.

While the energy flow coincides with the local density of linear momentum of the electromagnetic field, the local density of angular momentum is more difficult to discuss, essentially because the necessity of an intrinsic component of the angular momentum related to the spin of the photon forbids the total angular momentum density to be simply defined as $\br\times\PI(\br)$ \cite{bliokh2010angular}. This can only be done with the orbital part of the Poynting vector which thus directly gives the time-averaged density of orbital angular momentum as $\bLamb_O = \br \times \PI_O$. The dual-symmetry approach of Ref. \cite{bliokh2014conservation} however defines gauge invariant spin $\bLamb_S$ and orbital $\bLamb_O$ angular momentum time-averaged densities as:
\begin{align}
\bLamb_S &=  \frac{1}{k^2} \bPhi \label{SAM_terms} \\
\bLamb_O &=  -\frac{\omega}{4 k^2} \Im \left[\sum_j \eps E_j (\br \times \nabla) E_j^{*} +  \mu H_j (\br \times \nabla) H_j^{*} \right] \label{OAM_terms}
\end{align}
where $E_j$ and $H_j$ are the scalar components of the vector fields $\bE_0, \bH_0$. One can check that the latter expression for the orbital angular momentum fulfills $\bLamb_O = \br \times \PI_O$, with the dual-symmetric orbital linear momentum given in Eq.~(\ref{PI_O_dual}). Let us insist once again that the total angular momentum $\bLamb=\bLamb_O + \bLamb_S$ that arises from the definitions (\ref{SAM_terms}-\ref{OAM_terms}) does not necessarily verify $\bLamb=\br \times \PI$, as the spin part of the angular momentum has been introduced in a different way via magnetic and electric vector potentials \cite{bliokh2014conservation}.

\subsection{Optical forces and torques on a chiral dipole}

 We summarize here a few results derived earlier in Ref. \cite{canaguier2013mechanical}. The electric and magnetic dipolar moments $\bcalP=\Re[\bp_0(\br) e^{-\imath \omega t}]$ and $\bcalM=\Re[\bbm_0(\br) e^{-\imath \omega t}]$ of a chiral dipole depend on both the incident electric and magnetic fields:
\begin{align}\label{def_dipole_alpha_beta_chi}
\left( \begin{array}{c} \bp_0 \\ \bbm_0 \end{array} \right) =  \left( \begin{array}{cc} \alpha \eps_m &   \imath \chi \sqrt{\eps_m \mu_m}   \\ -\imath \chi   \sqrt{\eps_m \mu_m} & \beta \mu_m \end{array} \right) \times  \left( \begin{array}{c}\bE_0 \\\bH_0 \end{array} \right)
\end{align}
where the electric, magnetic and mixed electric-magnetic dipole polarizabilities $\alpha, \beta, \chi$ have the dimension of a volume. For simplicity, we assume bi-isotropic response such that $(\alpha,\beta,\chi)$ are complex numbers. The time-averaged optical force applied to such a chiral dipole by a general harmonic field splits into additive non-chiral and chiral contributions. Each contribution is separable into reactive and dissipative components that respectively involve the real and imaginary parts of the $\alpha,\beta$ or $\chi$ polarizabilities \cite{canaguier2013mechanical}:
\begin{align}
\bF^\mathrm{reac}_{\alpha,\beta} &= \Re[\alpha] \nabla W_E + \Re[\beta] \nabla W_H \label{F_achi_r} \\
 \bF^\mathrm{diss}_{\alpha,\beta} &=  \Im[\alpha] \frac{k^2}{\omega} \PI_O^{(E)} +  \Im[\beta] \frac{k^2}{\omega} \PI_O^{(H)}  \label{F_achi_d} \\
  \bF^\mathrm{reac}_\chi &= \Re[\chi] \frac{1}{k} \nabla K  \label{F_chi_r} \\
 \bF^\mathrm{diss}_\chi &=  \Im[\chi] \frac{2k}{\omega} \left( \bPhi - \frac{\nabla \times \PI }{2} \right) \label{F_chi_d} ~ .
\end{align}
The same non-chiral and chiral separation can be done on the time-averaged optical torque applied to the dipole with:
\begin{align}
\bN_{\alpha,\beta} &=  \Im[\alpha] \eps_m \bPhi_E + \Im[\beta] \mu_m \bPhi_H \label{N_achi} \\
\bN_\chi &=  \Im[\chi] \frac{2k}{\omega} \PI \label{N_chi}.
\end{align}

There is a remarkable symmetry in these expressions that directly stems from the fact that the chiral content of the harmonic field acts on the dipole through the real and imaginary parts of its chiral polarizability $\chi$. This is further emphasized when noting that the dissipative components of the force and the curl of the corresponding torque can lead to simple closed relations
\begin{align}
 \bF^\mathrm{diss}_{\alpha,\beta} + \frac{\nabla \times \bN_{\alpha,\beta}}{2} &= \Im[\alpha+\beta] \frac{k^2}{\omega} \PI \label{relation_achi} \\
  \bF^\mathrm{diss}_\chi + \frac{\nabla \times \bN_\chi}{2} &= \Im[\chi] \frac{2k}{\omega} \bPhi \label{relation_chi}
\end{align}
that reinforce the connection between optical energy and non-chiral forces on the one hand and between optical chirality and chiral forces, on the other hand \cite{canaguier2013mechanical}. 

\begin{figure}[htbp]
\centering{
\vspace{0.2cm}
\includegraphics[width=0.45\textwidth]{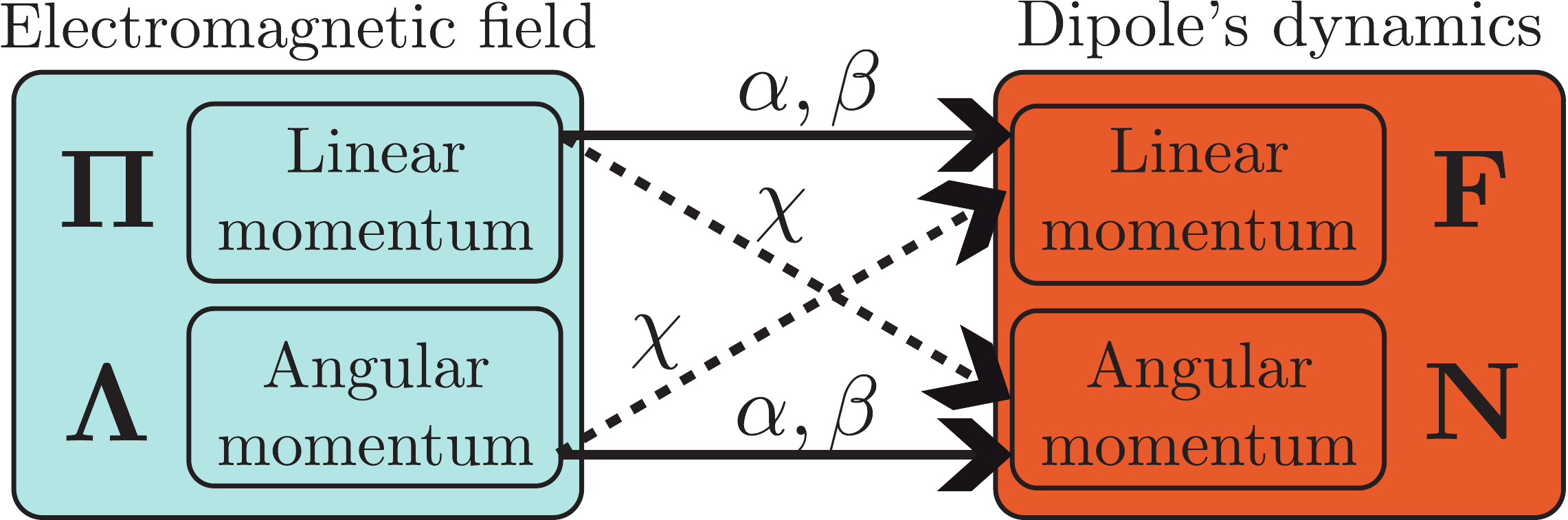}
\vspace{0.cm}
}
\caption{Schematics of the ``direct'' and ``crossed'' momentum transfers interpretation of optical forces and torques applied to a chiral dipole. The non-chiral ($\alpha, \beta$) component of the dissipative force (resp. of the torque) couples linear (resp. angular) momentum of the light to linear (resp. angular) momentum of the particle, while the chiral ($\chi$) component of the dissipative force and of the torque cross-couples linear to angular momenta in both directions.}
 \label{fig:tableau}
\end{figure}

These relations clearly show how the mechanical actions exerted by the optical field on the dipole are connected with light-matter momentum transfers. The dissipative part of the non-chiral force $ \bF^\mathrm{diss}_{\alpha,\beta}$ can be interpreted, according to Eq.~(\ref{F_achi_d}), as a transfer of orbital linear momentum of light to linear momentum for the dipole. The torque $\bN_{\alpha,\beta}$ given in Eq.~(\ref{N_achi}) corresponds to a transfer of spin angular momentum of light to angular momentum for the dipole. Involving the polarizabilities $(\alpha,\beta)$, these linear-to-linear and angular-to-angular momentum transfers are sketched in Fig.~\ref{fig:tableau} (horizontal arrows). 
In contrast, chiral dynamical effects cross-couple linear and angular momenta between the field and the dipole through the chiral mixed polarizability $\chi$. Indeed, the linear momentum of light is coupled to the dipole angular momentum at the level of the chiral torque $\bN_\chi$ given in Eq.~(\ref{N_chi}), when the chiral dissipative force $\bF^\mathrm{diss}_\chi$ given in Eq.~(\ref{F_chi_d}) corresponding to linear momentum for the dipole has its optical source in the angular momentum of light. These $\chi$-based crossed momentum transfers are depicted in Fig.~\ref{fig:tableau} (dashed-arrows).

As we will now discuss in detail, pulling optical forces and left-handed torques stem from a competition between these two types (direct and crossed) of momentum transfers that determine the balance between the non-chiral and the chiral dynamical action of light on the dipole. This balance eventually fixes the actual sign of the resulting optical forces and torques. 

%---------------------
\section{Negative optical forces in the dipolar regime}
%---------------------

Such competing momentum transfers are most easily discussed by considering a single right-handed circularly polarized plane wave (CPL), propagating in the $z>0$ direction:
\begin{align*}
&\bE_0 = \frac{E_0}{\sqrt{2}} ~ e^{\imath k z} \left( \begin{array}{c} 1 \\ \imath \\ 0 \end{array}\right) ~ , 
&\bH_0 = \frac{H_0}{\sqrt{2}}~ e^{\imath k z} \left( \begin{array}{c} -\imath \\ 1 \\ 0 \end{array}\right) 
\end{align*}
where $H_0=E_0 \sqrt{\eps_m/\mu_m}$. Introducing the field intensity $I_0 = E_0 H_0^*$, the Poynting vector $\PI=\frac{I_0}{2}\hat{\bz}$ gives the optical linear momentum and the chirality flux $\bPhi = \frac{k I_0}{2} \hat{\bz}$ the spin angular momentum. Both are homogeneous and pointing in the $z>0$ direction. As a consequence of the homogeneity, the spin part of the Poynting vector vanishes and $\PI=\PI_O$ is valid for any of the three spin-orbit decompositions defined in Eqs.~(\ref{PI_decompo},\ref{PI_O_dual}) \cite{canaguier2013force}. The optical angular momentum has an intrinsic spin component $\bLamb_S = \frac{I_0}{2k} \hat{\bz} $ in the longitudinal direction and an origin-dependent orbital component $\bLamb_O = \frac{I_0}{2} \left( y \hat{\bx} -x \hat{\by} \right)$ in the transverse plane. 

\subsection{Dipolar chiral particle model}

For a spherical nanoparticle of radius $R$, assumed to be small compared to spatial variations of the field (rather than smaller than the wavelength), the dipolar $(\alpha,\beta,\chi)$ polarizabilities can be derived from the macroscopic susceptibilities of the bulk matter constituting the nanoparticle, with electric permittivity $\eps$, magnetic permeability $\mu$ and chirality parameter $\kappa$. While $(\eps,\mu)$ give the nanoparticle refractive index $n$, the $\kappa$ parameter characterizes its chirality. It can be interpreted as the difference of the effective refractive index for left and right-handed circularly polarized wave (see section \ref{section:chiral_medium} in the Appendix for details). 

In the quasi-static limit (see Appendix for details), polarizabilities and susceptibilities are connected, with:
\begin{align}
\alpha &= 4\pi R^3 \frac{(\eps_r-1)(\mu_r+2) - \kappa^2}{(\eps_r+2)(\mu_r+2) - \kappa^2} \label{def_alpha} \\
\beta &= 4\pi R^3 \frac{(\mu_r-1)(\eps_r+2) - \kappa^2}{(\eps_r+2)(\mu_r+2) - \kappa^2} \label{def_beta} \\
\chi &= 12\pi R^3 \frac{ \kappa}{(\eps_r+2)(\mu_r+2) - \kappa^2} \label{def_chi}
\end{align}
where $\eps_r=\eps/\eps_m$ and $\mu_r=\mu/\mu_m$ are the relative permittivity and permeability, respectively. For $\kappa=0$, the usual forms of the Clausius-Mossotti polarizabilities $(\alpha, \beta)$ are recovered \cite{serdiukov2001electromagnetics}. 

It is important to note that the chiral parameter $\kappa$ is not related to $\chi$ only, but is also involved in the purely electric $\alpha$ and magnetic $\beta$ polarizabilities. Unexpectedly, this leads to the interesting fact that even so-called ``non-chiral'' optical forces that do not depend on $\chi$ can still depend on the chiral properties of the illuminated object through the modification of $\alpha$ and $\beta$ induced by $\kappa\neq 0$ in  Eqs.~(\ref{def_alpha},\ref{def_beta}). Nevertheless, this modification in $\alpha$ and $\beta$ depends on $\kappa^2$, meaning that two enantiomers characterized by $\pm \kappa$ will experience the same influence from the light field. From the separation of contributions given above -Eqs. (\ref{F_achi_r},\ref{F_achi_d},\ref{F_chi_r},\ref{F_chi_d})- the notion of ``non-chiral'' optical forces is to be understood with respect to this identity of action between two enantiomers, rather than with respect to the more stringent definition of a force field defined in the strict $\kappa\rightarrow 0$ limit.

Also, the denominator in Eq. (\ref{def_chi}) prevents a one-to-one relation between the real and imaginary parts of $\chi$ and $\kappa$, with $\Re [\kappa]$ corresponding to the rotatory power and $\Im [\kappa]$ the circular dichroism of material composing the object. Indeed, when for instance $\eps$ or $\mu$ has a non zero imaginary parts, the real and imaginary parts of $\kappa$ are mixed to yield a complex $\chi$. This yields the important conclusion that matter with properties of either rotatory powers or circular dichroism can both be used to generate either dissipative and reactive chiral optical forces. 

\subsection{Pulling force}

The pulling effect corresponds to the displacement of the illuminated particle in a direction opposite to the propagation direction of the illumination beam \cite{novitsky2011single,sukhov2011negative,chen2011optical}. Because it measures the flux of the energy contained within the beam, the optical linear momentum is the most natural way of defining the propagation direction of a beam. In this section we show that the influence of chirality with respect to optical forces opens a new route towards pulling forces where strong negative force components are created by crossed transfers from optical angular momenta to linear momenta of the particle. 

With the chosen CPL configuration, the energy and chirality densities $W,K$ of the optical field are homogeneous. Therefore, the optical force reduces to its dissipative non-chiral and chiral components (\ref{F_achi_d},\ref{F_chi_d}) 
\begin{align}\label{forces_CPL}
&F_{\alpha,\beta}^\mathrm{diss} = \frac{\omega I_0}{2k^2} \Im[\alpha+\beta] \hat{\bz} ~ , 
&F_\chi^\mathrm{diss} = \frac{\omega I_0}{k^2} \Im[\chi] \hat{\bz} 
\end{align}
that add up as the total force exerted on the dipole. Each component is associated with a specific momentum transfer. The non-chiral component comes from optical linear momentum $\PI_O=\PI$, mediated by the diagonal terms $(\alpha, \beta)$ of the polarizability matrix given in Eq. (\ref{def_dipole_alpha_beta_chi}). Being proportional to the poynting vector, this component yields a positive force $F_{\alpha,\beta}^\mathrm{diss}$ since $\Im[\alpha+\beta]\geq0$ for passive materials. The chiral component comes from spin angular linear momentum $\bLamb_S$, mediated by the chiral mixed polarizability $\chi$. Contrasting with the non-chiral component, there is no restriction on the  sign of $F_\chi^\mathrm{diss}$ since the sign of $\chi$ is reversed for opposite enantiomers. As a consequence, if the crossed momentum transfer associated with $\chi$ is strong enough, a negative chiral force could overcome the positive non-chiral force and give a total pulling force, directed in the $(z<0)$ direction. From Eq.~(\ref{forces_CPL}), this can happen if and only if $\chi$ has a negative imaginary part large enough for $\Im \left[ (\alpha+\beta)/2 + \chi \right]$ to  become negative.

When modeling the optical response of the particle, we choose $\eps/\eps_0=-7.837 + 1.155 \imath$, $\mu=\mu_0$, i.e. bulk gold at a wavelength $\lambda = \frac{2\pi c}{\omega} = 594$ nm in vacuum. The chirality parameter $\kappa$ becomes the free parameter that we take purely imaginary for simplicity. With a refractive index $n=\sqrt{\eps/\eps_0}=0.206 + 2.807 \imath$, the chirality parameter is bounded with $|\kappa |\leq 2.807$ in order to ensure that the particle is made from a passive material. The surrounding medium is taken as water, with $\eps_m/\eps_0=(1.33)^2$ and $\mu_m=\mu_0$. 

\begin{figure}[htbp]
\centering{
\includegraphics[width=0.45\textwidth]{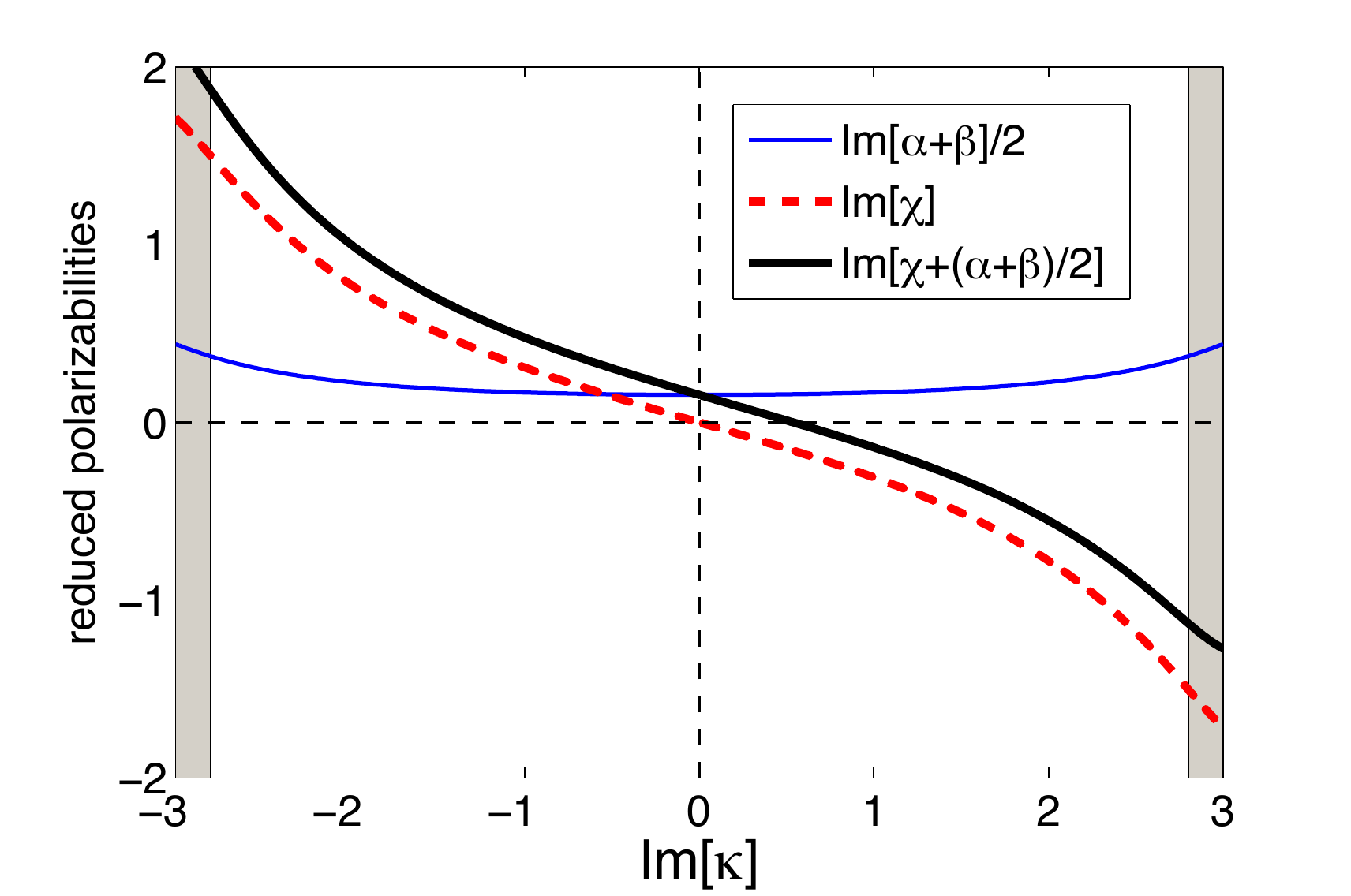}}
\caption{(colors online) Imaginary parts of the polarizabilities, normalized by $4\pi R^3$, as functions of the imaginary parameter $\kappa$ characterizing the chirality of the dipolar particle. The non-chiral component of the optical force is given by $\Im[\alpha+\beta]/2$ (blue thin curve), which is an even function of $\kappa$. The chiral component of the force is given by $\Im[\chi]$ (red dashed-curve), which is an odd function of $\kappa$. The total optical force is given by the sum of the two terms (dark solid curve). For $\Im[\kappa]\gtrsim 0.54$, the chiral component of the force is strong enough to reverse the direction of the total force experienced by the dipolar particle. Grey vertical areas bound the values for $\Im[\kappa]$ in relation with passivity.}
 \label{fig:pola_wrt_kappa}
\end{figure}

Within the allowed $\kappa$ values, we plot the imaginary parts of the three polarizabilities in Fig.~\ref{fig:pola_wrt_kappa}. This gives the opportunity to look at the balance between the forces given in Eq.~(\ref{forces_CPL}). As expected, because $\Im[\alpha+\beta] >0$, the non-chiral force $F_{\alpha,\beta}^\mathrm{diss}$ will always be positive, pouting in the direction of the Poynting vector. 
As discussed above, the non-chiral force component actually depends on the value of the chirality parameter. This dependency clearly shows up in  Fig.~\ref{fig:pola_wrt_kappa} where the even character of $\Im[\alpha+\beta]$ reflects the appearance of $\kappa^2$ in the expressions (\ref{def_alpha},\ref{def_beta}) of $\alpha$ and $\beta$. Then, the chiral force component $F_\chi^\mathrm{diss}$, which is given by $\Im[\chi]$, is an odd function of $\kappa$ and can hence lead to positive as well as negative forces. 

Finally, the total force, proportional in the CPL configuration to $\Im \left[ (\alpha+\beta)/2 + \chi \right]$, is plotted as the dark plain curve in Fig.~\ref{fig:pola_wrt_kappa}. Interestingly, for values of $\kappa$ within the allowed range, the amplitude of the negative chiral force can turn large enough to overcome the positive non-chiral force. For $\Im[\kappa] \gtrsim 0.54$ indeed, the total optical force becomes negative. As a spectacular consequence, a particle made of a material with such a chiral parameter will experience a pulling force when illuminated by a simple CPL. Again, this effect is based on crossed momentum transfers induced by chirality as sketched in Fig.~\ref{fig:tableau}. Such crossed transfers can cancel and overcome the usual momentum transfer for a particle sufficiently chiral.

\subsection{Left-handed torque}

The second kind of negative mechanical effect is the left-handed optical torque, where the electromagnetic field exerts a torque on a particle in a direction opposite to its angular momentum \cite{hakobyan2014left}. In this section, we show that cross-transfers from the optical linear momentum to the particle angular momentum lead to situations where such negative torques arise.  

In the simple CPL configuration, we consider the optical torque applied to the dipolar particle, whose non-chiral and chiral components are 
\begin{align}\label{torques_CPL}
&N_{\alpha,\beta} = \frac{k I_0}{2\omega} \Im[\alpha+\beta] \hat{\bz} ~ , 
&N_\chi= \frac{k I_0}{\omega} \Im[\chi] \hat{\bz} ~ .
\end{align}
Just as discussed for the force components, we stress that the non-chiral torque has a fixed signed with $\Im[\alpha+\beta] \geq 0$ while the chiral torque is reversed when considering opposite enantiomers. The total torque is determined by the value of $\Im[(\alpha+\beta)/2 +\chi ]$ which implies that the discussion made above on the dependence of the different polarizabilities on $\kappa$ plotted Fig.~\ref{fig:pola_wrt_kappa} also apply for the torque. As a consequence, the total optical torque can become negative for a sufficiently chiral medium. This left-handed torque, heading in the direction opposite to the field angular momentum, can be easily interpreted in the context of momentum transfers presented in Fig.~\ref{fig:tableau}: for strongly chiral particle, the crossed transfers, coming from optical linear momentum $\PI$ can overcome the usual transfer from optical angular momentum $\bLamb$. 

These two examples show that chiral optical forces and torques derived in the dipolar regime are readily able to overcome the usual direct momentum transfers, revealing the potential of chirality for generating pulling optical forces and left-handed optical torques. For clarity, we have chosen here a simple optical field: a single propagative plane wave, circularly polarized. But obviously, momentum transfers mediated by chirality can be engineered with more complex field, such as focussed beam or vortex beam carrying orbital angular momentum.

It is interesting to emphasize that such negative dynamical phenomena have been obtained in the strict dipolar regime, that does not include self-interaction mechanical effects such as generated by the scattered field. When the scattering is included in the calculation, one must also consider beyond-dipolar terms in the optical response of the particle, as they can be on the same order as self-interaction contributions. We thus consider below the multipolar expansion of the scattering by a chiral sphere and derive chiral forces and torques accounting for both the field scattered by the particle and the higher-order multipoles in the optical response of the particle.

%---------------------
\section{Multipolar effects on chiral optical forces}
%---------------------

It is interesting to extend the discussion on the interplay of chirality with optical forces beyond the dipolar case by considering optical forces applied to chiral sphere of arbitrary sizes. This can be done by expanding the known multipolar calculations of optical forces (see \cite{canaguier2014transverse} and references therein) to the case of materials with chiral properties. This expansion gives rise to a particularly rich landscape of results, where chirality becomes intertwined with multipolar effects. 

We will focus on two simple situations where direct (linear-to-linear and angular-to-angular) momentum transfers, represented as horizontal lines in Fig.~\ref{fig:tableau}, are cancelled. In such situations, optical forces and torques are only determined from the remaining crossed momentum transfers. We show here that such transfers can be strongly modified due to multipolar resonances and that this modification causes sign flips for the optical force and torque when the size of the sphere increases. This break of symmetry is a rather surprising effect of the interplay between chirality and multipolar effects for finite-size objects.

\subsection{Multipolar calculation of the optical force and torque on a chiral sphere}

We first summarize the main results related to the scattering of a chiral sphere, treated for instance in \cite{guzatov2011chiral,wu2012calculation,shang2013analysis,shang2013rainbow}. We give here explicit expressions that are in agreement with our choice of conventions. The coefficients $A_{\ell,m}^{S}, B_{\ell,m}^{S}$ of the scattered field in the basis of spherical modes can be obtained from the coefficients  $A_{\ell,m}, B_{\ell,m}$ of the incident field (see \cite{canaguier2014transverse} for a derivation of the latter coefficients) with additional non-diagonal terms:
\begin{align}\label{AB_sca}
\left( \begin{array}{c} A_{\ell,m}^{S} \\ B_{\ell,m}^{S} \end{array} \right) 
= \left( \begin{array}{cc} a_\ell  & \imath c_\ell \\ -\imath c_\ell & b_\ell \end{array} \right) 
\times \left( \begin{array}{c} A_{\ell,m} \\ B_{\ell,m} \end{array} \right)  ~ .
\end{align}
The three multipolar coefficients $a_\ell, b_\ell, c_\ell$ are generalized Mie coefficients for chiral media. For a sphere made of a material with permittivity $\eps$, permeability $\mu$ and chiral parameter $\kappa$, it is useful to introduce the three adimensional quantities:
\begin{align}\label{def_x1_x2}
& x_0 = \frac{\omega n_m R}{c} &\nonumber \\
& x_1= \frac{\omega (n+\kappa) R}{c} & x_2= \frac{\omega (n-\kappa) R}{c} ~ .
\end{align}
The generalization of the Mie coefficients for a chiral sphere can then be nicely casted in a compact form:
\begin{align}\label{Mie_coefficients}
a_\ell  &= -K_0 \frac{R_1+R_2}{S_1+S_2} \nonumber \\
b_\ell  &= -K_0 \frac{P_1+P_2}{Q_1+Q_2} \nonumber \\
c_\ell  &= -K_0 \frac{T_1-T_2}{Q_1+Q_2}
\end{align}
where we define: 
\begin{align*}
K_0 &= \frac{\psi_\ell(x_0)}{\xi_\ell (x_0)} = \frac{J_{\ell+1/2}(x_0)}{H^{(1)}_{\ell+1/2}(x_0)} \nonumber \\
P_j &= \frac{D_\ell^{(1)} (x_j) - \eta D_\ell^{(1)}(x_0)}{\eta D_\ell^{(1)}(x_j)-D_\ell^{(3)}(x_0)}  ~ ~ , ~ ~ ~ j \in \{1, 2 \}\nonumber \\
Q_j &= \frac{\eta D_\ell^{(3)} (x_0) - D_\ell^{(1)}(x_j)}{\eta D_\ell^{(1)}(x_j)-D_\ell^{(3)}(x_0)}  ~ ~ , ~ ~ ~ j \in \{1, 2 \}\nonumber \\
R_j &= \frac{\eta D_\ell^{(1)} (x_j) - D_\ell^{(1)}(x_0)}{D_\ell^{(1)}(x_j) - \eta D_\ell^{(3)}(x_0)}  ~ ~ , ~ ~ ~ j \in \{1, 2 \}\nonumber \\
S_j &= \frac{D_\ell^{(3)}(x_0) - \eta D_\ell^{(1)}(x_j)}{D_\ell^{(1)}(x_j) - \eta D_\ell^{(3)}(x_0)}  ~ ~ , ~ ~ ~ j \in \{1, 2 \}\nonumber \\
T_j &= \frac{\eta D_\ell^{(1)}(x_j) - D_\ell^{(1)}(x_0)}{\eta D_\ell^{(1)}(x_j) - D_\ell^{(3)}(x_0)}  ~ ~ , ~ ~ ~ j \in \{1, 2 \}\nonumber 
\end{align*}
and
\begin{align}\label{D1_D3}
D_\ell^{(1)}(x) &= \frac{\psi_\ell^{'}(x)}{\psi_\ell(x)} = \frac{x J_{\ell-1/2}(x)- \ell J_{\ell+1/2}(x)}{x J_{\ell+1/2}(x)} \nonumber \\
D_\ell^{(3)}(x) &= \frac{\xi_\ell^{'}(x)}{\xi_\ell(x)} = \frac{x H^{(1)}_{\ell-1/2}(x)- \ell H^{(1)}_{\ell+1/2}(x)}{x H^{(1)}_{\ell+1/2}(x)} ~ .
\end{align}
Of course, for a vanishing chirality parameter $\kappa$, one has $x_1=x_2$, $c_\ell=0$ and the usual Mie coefficients are recovered for $a_\ell$ and $b_\ell$.

The force and torque applied to the sphere can then be calculated by generalizing the formula obtained in the non-chiral case -see \cite{canaguier2014transverse}. In the present expressions, we have replaced the explicit Mie coefficients with the obtained scattered fields given in Eq.~(\ref{AB_sca}):
\begin{widetext}
\begin{align}\label{Fxy_multipolar}
\frac{F_x + \imath F_y}{n_m I_0 / c} = &\frac{\imath}{4} x_0^2 R^2  \sum_{\ell=1}^{\infty} \sum_{m=-\ell}^{\ell} \left\{ \sqrt{\frac{(\ell+m+2)(\ell+m+1)}{(2\ell+1)(2\ell+3)}} \ell (\ell+2) \right.   \nonumber \\
& ~ ~ ~ ~ ~ ~ \times \left[ 2 A_{\ell,m}^{S} A_{\ell+1,m+1}^{S~*} - A_{\ell,m}^S A_{\ell+1,m+1}^{*} - A_{\ell,m} A_{\ell+1,m+1}^{S~*} \right.+ \left.  2 B_{\ell,m}^{S} B_{\ell+1,m+1}^{S~*} - B_{\ell,m}^S B_{\ell+1,m+1}^{*} - B_{\ell,m} B_{\ell+1,m+1}^{S~*} \right] \nonumber \\
&+\sqrt{\frac{(\ell-m+2)(\ell-m+1)}{(2\ell+1)(2\ell+3)}} \ell (\ell+2)  \nonumber \\
& ~ ~ ~ ~ ~ ~  \times \left[ 2 A_{\ell+1,m-1}^{S} A_{\ell,m}^{S~*} - A_{\ell+1,m-1}^S A_{\ell,m}^{*} - A_{\ell+1,m-1} A_{\ell,m}^{S~*} \right.+ \left.  2 B_{\ell+1,m-1}^{S} B_{\ell,m}^{S~*} - B_{\ell+1,m-1}^S B_{\ell,m}^{*} - B_{\ell+1,m-1} B_{\ell,m}^{S~*} \right] \nonumber \\
& + \sqrt{(\ell+m+1)(\ell-m)}  \nonumber \\
& ~ ~ ~ ~ ~ ~ \times  \left[ 2 A_{\ell,m}^{S} B_{\ell,m+1}^{S~*} - A_{\ell,m}^S B_{\ell,m+1}^{*} - A_{\ell,m} B_{\ell,m+1}^{S~*} \right. -  \left.  2 B_{\ell,m}^{S} A_{\ell,m+1}^{S~*} - B_{\ell,m}^S A_{\ell,m+1}^{*} - B_{\ell,m} A_{\ell,m+1}^{S~*} \right] \left. \vphantom{\sqrt{\frac{(\ell+m+2)(\ell+m+1)}{(2\ell+1)(2\ell+3)}}} \right\} 
\end{align}
\begin{align}\label{Fz_multipolar}
\frac{F_z}{n_m I_0 / c} =& - \frac{x_0^2 R^2}{2} \sum_{\ell=1}^{\infty} \sum_{m=-\ell}^{\ell} \left\{ \vphantom{\sqrt{\frac{(\ell+m+2)(\ell+m+1)}{(2\ell+1)(2\ell+3)}}} \sqrt{\frac{(\ell-m+1)(\ell+m+1)}{(2\ell+1)(2\ell+3)}} \ell (\ell+2) \right. \nonumber \\
& ~ ~ ~ ~ ~ ~  \times  \Im \left[ 2 A_{\ell+1,m}^S A_{\ell,m}^{S~*} - A_{\ell+1,m}^S A_{\ell,m}^{*} - A_{\ell+1,m} A_{\ell,m}^{S~*} \right.    + \left.  2 B_{\ell+1,m}^S B_{\ell,m}^{S~*} - B_{\ell+1,m}^S B_{\ell,m}^{*} - B_{\ell+1,m} B_{\ell,m}^{S~*} \right]  \nonumber \\
& + m \Im \left[  2 A_{\ell,m}^S B_{\ell,m}^{S~*} - A_{\ell,m}^S B_{\ell,m}^{*} - A_{\ell,m} B_{\ell,m}^{S~*} \right]  \left. \vphantom{\sqrt{\frac{(\ell+m+2)(\ell+m+1)}{(2\ell+1)(2\ell+3)}}} \right\} 
\end{align}

\begin{align}\label{Nxy_multipolar}
\frac{N_x + \imath N_y}{n_m I_0 / c} = & - \frac{x_0 R^3}{4}  \sum_{\ell=1}^{\infty} \sum_{m=-\ell}^{\ell} \ell (\ell+1) \sqrt{(\ell-m)(\ell+m+1)} \nonumber \\
& ~ ~ ~  \times \left[ 2 A_{\ell,m}^S A_{\ell,m+1}^{S~*} - A_{\ell,m}^S A_{\ell,m+1}^{*} - A_{\ell,m} A_{\ell,m+1}^{S~*} \right. 
 + \left. 2 B_{\ell,m}^S B_{\ell,m+1}^{S~*} - B_{\ell,m}^S B_{\ell,m+1}^{*} - B_{\ell,m} B_{\ell,m+1}^{S~*} \right]  
\end{align}
\begin{align}\label{Nz_multipolar}
\frac{N_z}{n_m I_0 / c} = & - \frac{x_0 R^3}{4}  \sum_{\ell=1}^{\infty} \sum_{m=-\ell}^{\ell} m (\ell+1)  \left[ 2 A_{\ell,m}^S A_{\ell,m}^{S~*} - A_{\ell,m}^S A_{\ell,m}^{*} - A_{\ell,m} A_{\ell,m}^{S~*} \right.  + \left. 2 B_{\ell,m}^S B_{\ell,m}^{S~*} - B_{\ell,m}^S B_{\ell,m}^{*} - B_{\ell,m} B_{\ell,m}^{S~*}  \right]  
\end{align}
\end{widetext}
where $n_m I_0 / c =  \eps_m |E_0|^2 = \mu_m |H_0|^2 $ has the dimension of a pressure. Note that the present expressions have a factor $2$ difference with respect to \cite{canaguier2014transverse} due to a different definition of $I_0$.

\subsection{Sign flips of the torque}

We first focus on the crossed momentum transfer from the linear momentum of light to the angular momentum of the particle, operated by the chirality-dependent part of the torque. In order to study this transfer and how it is affected by multipoles, we consider a single linearly polarized plane wave, propagating along the $z>0$ direction:
\begin{align*}
&\bE_0 = E_0 ~ e^{\imath k z} \left( \begin{array}{c} 1 \\ 0 \\ 0 \end{array}\right) ~ , 
&\bH_0 = H_0~ e^{\imath k z} \left( \begin{array}{c} 0 \\ 1 \\ 0 \end{array}\right). 
\end{align*}
In this optical configuration, the spin angular momentum of light vanishes ($\bLamb_S=\mathbf{0}$), while the orbital angular momentum $\bLamb_O = \frac{I_0}{2} (y \hat{\bx} - x \hat{\by})$ is in the transverse plane. It follows that $N_x=N_y=0$ so that $N_z$ is the only torque exerted on the particle in the propagation direction, stemming from the crossed momentum transfer we focus on. 

The derivation of the torque in the dipolar regime given in Eq.~(\ref{N_chi}) predicts a term $N_z$ originating from the Poynting vector $\PI$, and whose direction depends on the sign of $\Im[\chi]$. This implies an inversion of the direction of the torque under the change of enantiomer. Beyond the dipolar regime however, higher multipoles lead to a more complex influence of chirality on the optical torque.

This can be illustrated by considering a lossy dielectric sphere made of a chiral material with an optical response characterized by relative permittivity $\eps/\eps_0=(1.7+0.1 \imath)^2$, relative permeability $\mu/\mu_0=1$ and a small imaginary chiral parameter $\kappa = \pm 0.001 \imath$ that corresponds to 1$\%$ circular dichroism for the bulk material. 
\begin{figure}[htbp]
\centering{
\includegraphics[width=0.45\textwidth]{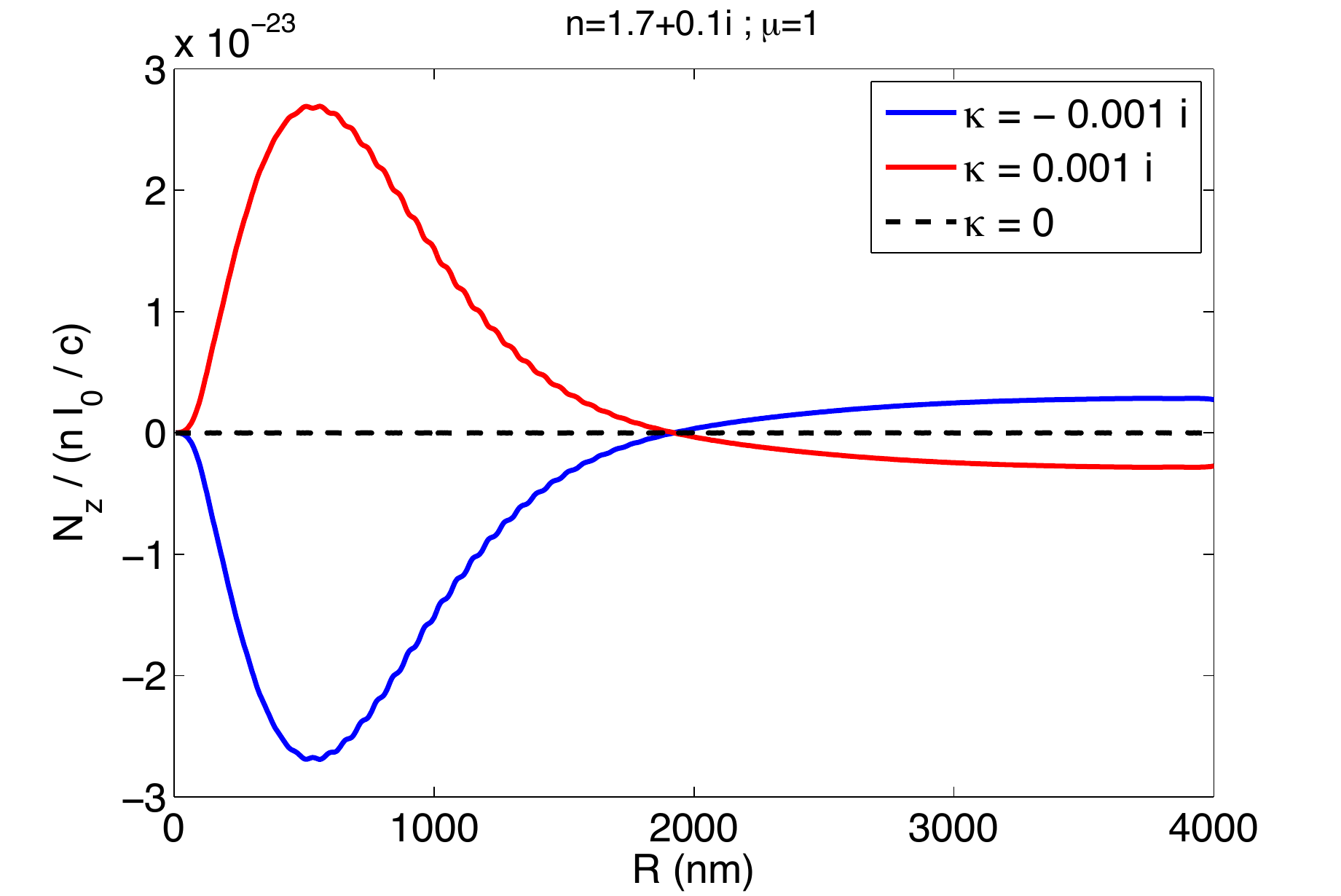}}
\caption{(colors online) Optical torque exerted by a propagative plane wave linearly polarized along the propagation direction, as a function of the sphere radius. The sphere is made of a dissipative dielectric, with $\sqrt{\eps/\eps_0} = n=1.7+0.1 \imath$, and characterized by a purely imaginary chirality parameter $\kappa=\pm0.001 \imath$. The numerical calculations include multipoles up to $\ell=60$.}
 \label{fig:torque_flip}
\end{figure}
The evolution of the optical torque $N_z$ exerted by the linearly polarized beam is shown in Fig.~\ref{fig:torque_flip} as a function of the sphere radius $R$ for opposite values of the chirality parameter $\kappa$. As expected, the torque vanishes for a non-chiral media $(\kappa=0)$ because in this case, neither optical linear nor angular momenta can be transferred to the particle. For small particles, a scaling $N_z \propto R^3 \Im[\kappa]$ appears, as expected from the dipolar result Eq.~(\ref{N_chi}). For larger spheres, the progressive contributions of higher multipoles to the total torque modify the whole size evolution so strongly that they are eventually responsible for a total sign flip of the torque. This sign flip happens for $R \gtrsim 2 \mu$m where the torque becomes positive (resp. negative) for negative (resp. positive) values of $\Im[\kappa]$. Note that for such large radii, the numerical calculations have to include spherical modes up to $\ell=60$ in order to get accurate estimations. 

\subsection{Sign flips of the force}

The complementary crossed momentum transfer can also be considered, namely the transfer from the angular momentum of light to the linear momentum of the particle in the form of a chiral optical force. In order to observe it, a configuration must be chosen in which the linear momentum of the incident light is zero. This is for instance obtained for the configuration of two incoherent counter-propagating CPL with opposite handedness:
\begin{align*}
&\bE_0 = E_0 ~ \cos kz  \left( \begin{array}{c} 1 \\ \imath \\ 0 \end{array}\right) ~ , 
&\bH_0 = H_0~ \sin k z \left( \begin{array}{c} 1 \\ \imath \\ 0 \end{array}\right) 
\end{align*}
which was extensively studied theoretically in the dipolar regime \cite{canaguier2013mechanical,cameron2014discriminatory} as well as experimentally \cite{tkachenko2014optofluidic} for its ability to mechanically separate enantiomers. Indeed, while the linear momenta of the two counter-propagating waves cancel each other ($\PI=0$), the spin angular momenta add up with a positive value $\bLamb=\frac{I_0}{2k} \hat{\bz}$ in the $z$ direction. The optical force is then only due to the sphere chirality which couples the angular momentum of light to the linear momentum of the sphere. 
\begin{figure}[htbp]
\centering{
\includegraphics[width=0.45\textwidth]{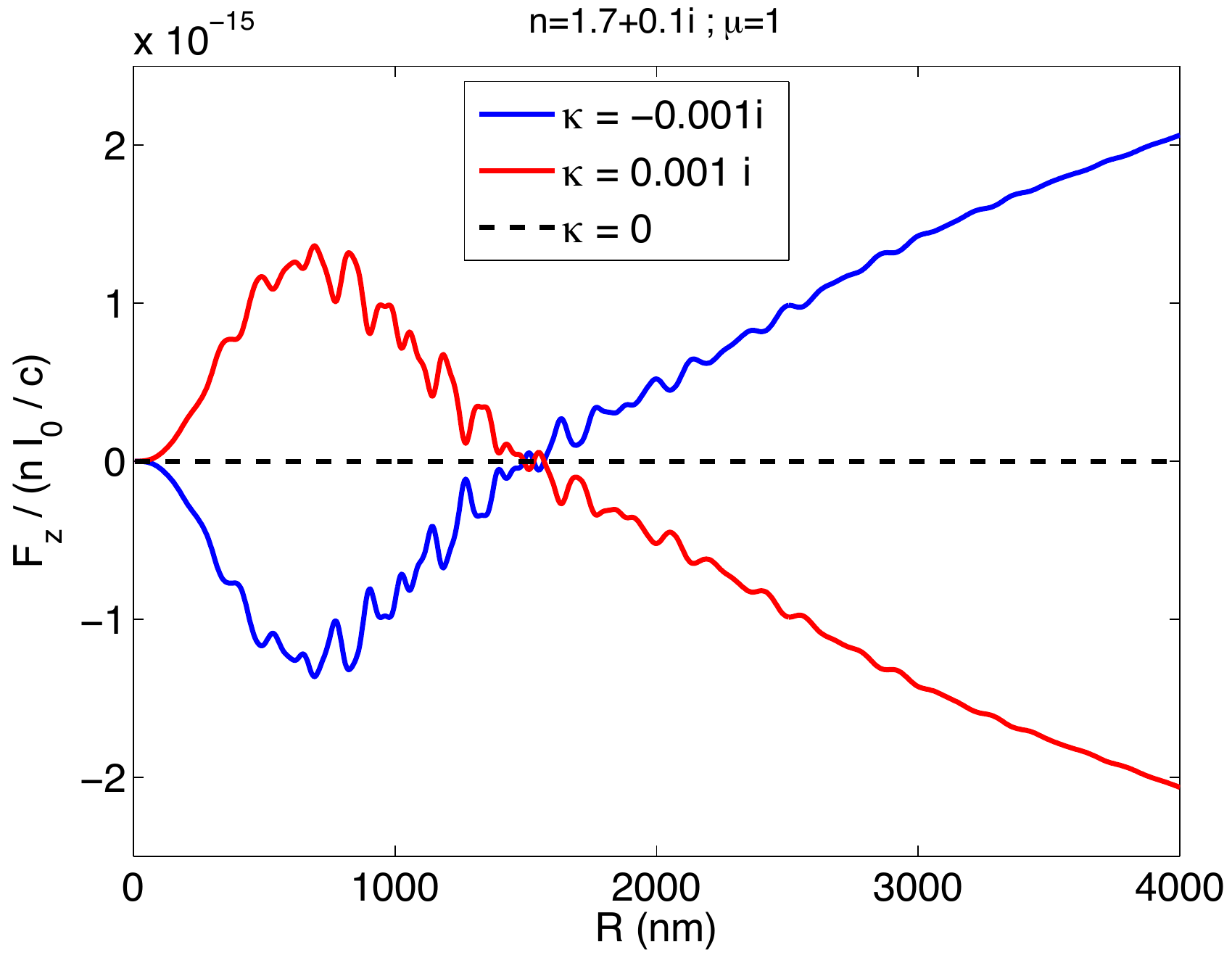}}
\caption{(colors online) Optical force exerted by two counter-propagating circularly polarized waves, as a function of the sphere radius. The two beams are incoherent and have opposite handedness, so that the $z$-component of the total ellipticity is positive. The sphere is made of a dissipative dielectric, with $\sqrt{\eps/\eps_0} = n=1.7+0.1 \imath$, and characterized by a purely imaginary chirality parameter $\kappa=\pm0.001 \imath$. The numerical calculation includes multipoles up to $\ell=70$.}
 \label{fig:force_flip}
\end{figure}
In such a configuration, the force, pointing along the $z$ direction, can be evaluated. The results are shown in Fig.~\ref{fig:force_flip} as a function of the sphere radius, for opposite values of the chirality parameter $\kappa$. We again check that the force drops to zero for a vanishing $\kappa$. The small sphere limit, where the force is positive for $\kappa = 0.001 \imath$, recovers the dipolar prediction, where the dipolar force given in Eq.~(\ref{F_chi_d}) has a positive (resp. negative) $z$-component for a positive (resp. negative) imaginary part of $\chi$. However, for a sphere of radius larger than 1.5 $\mu$m, the two curves crosses each other and the force in the $z$ direction becomes negative (resp. positive) for $\kappa = 0.001 \imath$ (resp. $\kappa = -0.001 \imath$). Here again, we have a situation where an increased size of the sphere results in a flip of the direction of the force applied to it. 

\section*{Conclusion}

We have shown how chirality leads to couple linear and angular momenta between light and matter, in such a way that negative dynamical effects can become predominant with the prediction of pulling light forces and left-handed torques in the dipolar regime. Such effects are also predicted to happen in the context of large spherical chiral objects where crossed momentum transfers are modified by the interplay between chirality and multipolar resonances. This shows that, beyond the dipolar limit, the actual size of the involved chiral objects and the multipolar components of their optical response are important parameters of the problem and even determine the signs of optical forces and torques. At all scales therefore, our work clarifies the connections between chirality and light-matter momentum transfers and it identifies new mechanisms that can be exploited experimentally in order to observe and exploit such surprising all-optical dynamic phenomena.

\section*{Acknowledgements}

We thank R. Carminati, T.W. Ebbesen, and J. A. Hutchison for fruitful discussions. We acknowledge support from the French program Investissement d'Avenir (Equipex Union).

\appendix

\section{Derivation of the dipolar polarizabilities for a small chiral sphere}

\subsection{Description of a chiral medium}

\label{section:chiral_medium}

We consider a bi-isotropic and causal medium with chiral properties (``Pasteur medium''), which is defined by its electric permittivity $\eps(\omega)$, magnetic permeability $\mu(\omega)$ and a chirality parameter $\kappa$, that are all three scalar, dimensionless and frequency-dependent quantities. The constitutive relations that connects the complex displacement field $\bD^\tot_0(\br)$ and the magnetic field $\bB^\tot_0(\br)$ to the electric field $\bE^\tot_0(\br)$ and magnetization field $\bH^\tot_0(\br)$ for the total electromagnetic field read:
\begin{align}\label{constitutive_DB}
\left( \begin{array}{c} \bD^\tot_0 (\br) \\ \bB^\tot_0 (\br) \end{array} \right) 
= \left( \begin{array}{cc} \eps & \imath \kappa / c \\ -\imath \kappa / c &  \mu \end{array} \right) 
\times 
\left( \begin{array}{c} \bE^\tot_0 (\br) \\ \bH^\tot_0 (\br) \end{array} \right) 
\end{align} 
where the $3\times3$ identity operators have been removed for clarity. When solving Maxwell equations in such a medium, one gets two kinds of waves, that are left and right-handed circularly polarized, associated with effective refractive indices $n_\pm = n \pm \kappa$. The complex chirality parameter $2\kappa$ is then a measure of the difference of the refractive index felt by two CPL with opposite handedness. The real part of $\kappa$ is thus associated with optical rotation, while its imaginary part drives circular dichroism.

 The relations (\ref{constitutive_DB}) can be re-written by defining the polarization density $\bP_0(\br) = \bD^\tot_0(\br) - \eps_0 \bE^\tot_0(\br)$ and the magnetization density $\bM_0(\br) = \bB^\tot_0(\br) - \mu_0 \bH^\tot_0(\br)$ of the medium:
\begin{align}\label{constitutive_PM}
\left( \begin{array}{c} \bP_0 (\br) \\ \bM_0 (\br) \end{array} \right) 
= \left( \begin{array}{cc} \eps - \eps_0 & \imath \kappa / c \\ -\imath \kappa / c & \mu -\mu_0 \end{array} \right) 
\times 
\left( \begin{array}{c} \bE^\tot_0 (\br) \\ \bH^\tot_0 (\br) \end{array} \right) ~ .
\end{align} 
This relation will be used together with Eq.~(\ref{def_dipole_alpha_beta_chi}), which involves the incident fields $\bE_0, \bH_0$ only, in order to get the dipolar polarizabilities $(\alpha, \beta, \chi)$ from the bulk parameter ($\eps, \mu, \kappa$). The only missing ingredient is the connection between incident and total fields.

\subsection{Lipmann-Schwinger equations}

Maxwell's equations for the electric and magnetic scattered fields, which are created by the sources $\bP_0, \bM_0$ in the particle, can be solved using the Green's dyadic function $\bG$ and its curl $\left[ \nabla \times \bG \right]$. These two dyadic functions allow to write the scattered fields as functions of its sources, and together with Eq.~(\ref{constitutive_PM}) lead to Lipmann-Schwinger relations between the incident and total electromagnetic fields:
\begin{widetext}
\begin{align}
\bE_0^\tot(\br) = \bE_0(\br) &+  \int_{V_P} \left[ k_0^2 (\eps/\eps_0-1) \bG(\br,\br',\omega) + k_0 \kappa \left[ \nabla \times \bG \right] (\br,\br',\omega) \right] \cdot \bE_0^\tot(\br') \d^3 \br' \nonumber \\
&+ \imath  \int_{V_P} \left[  k_0 \sqrt{\frac{\mu_0}{\eps_0}} (\mu/\mu_0-1) \left[ \nabla \times \bG \right] (\br,\br',\omega)+  k_0^2 \sqrt{\frac{\mu_0}{\eps_0}} \kappa \bG(\br,\br',\omega) \right] \cdot \bH_0^\tot(\br') \d^3 \br'  \label{Lipmann_E} \\
\bH_0^\tot(\br) = \bH_0(\br) &+  \int_{V_P} \left[ k_0^2 (\mu/\mu_0-1) \bG(\br,\br',\omega) + k_0 \kappa \left[ \nabla \times \bG \right] (\br,\br',\omega) \right] \cdot \bH_0^\tot(\br') \d^3 \br' \nonumber \\
&- \imath \int_{V_P} \left[  k_0 \sqrt{\frac{\eps_0}{\mu_0}} (\eps/\eps_0-1) \left[ \nabla \times \bG \right] (\br,\br',\omega)+  k_0^2 \sqrt{\frac{\eps_0}{\mu_0}} \kappa \bG(\br,\br',\omega) \right] \cdot \bE_0^\tot(\br') \d^3 \br' \label{Lipmann_H}
\end{align}
\end{widetext}
where $V_P$ is the volume of the particle.

\subsection{Dipolar and quasi-static limit}

In the dipolar limit, the electromagnetic fields are assumed to be homogeneous over the volume of the particle, which simplifies the integrals in Eqs.~(\ref{Lipmann_E},\ref{Lipmann_H}) as $\bE_0^\tot(\br')\simeq \bE_0^\tot(\br_P)$ and $\bH_0^\tot(\br') \simeq\bH_0^\tot(\br_P)$, where $\br_P$ is the position of the particle. These expressions can be further simplified in the quasi-static limit, where the integral of $\bG$ reduces to its singularity $-\frac{\Id}{3k_0^2}$ and the integral of $\left[ \nabla \times \bG \right]$ vanishes. This enables to express the total fields from the incident ones in a compact and simple expression:
\begin{widetext}
\begin{align}
\left( \begin{array}{c} \bE_0^\tot (\br_P) \\  \bH_0^\tot (\br_P) \end{array} \right)  = \frac{3}{(\eps/\eps_0+2)(\mu/\mu_0+2)-\kappa^2}
\left( \begin{array}{cc} \mu/\mu_0+2 & - \imath \sqrt{\frac{\mu_0}{\eps_0}} \kappa \\  \imath \sqrt{\frac{\eps_0}{\mu_0}} \kappa & \eps/\eps_0+2 \end{array} \right) 
 \cdot \left( \begin{array}{c} \bE_0 (\br_P) \\  \bH_0 (\br_P) \end{array} \right) ~ . 
\end{align}
\end{widetext}
 Together with Eq.~(\ref{constitutive_PM}), the above equations yield, from the incident fields, an expression of the polarization and magnetization densities in the particle. As these densities are homogeneous, they can be multiplied by the volume $V_P$ to get the electric and magnetic dipolar momenta of the particle. Then, by identification with the definition (\ref{def_dipole_alpha_beta_chi}) of the dipolar polarizabilities, one gets the generalization of the Clausius-Mossotti relations for a chiral particle:
 \begin{align*}
\alpha &= 4\pi R^3 \frac{(\eps_r-1)(\mu_r+2) - \kappa^2}{(\eps_r+2)(\mu_r+2) - \kappa^2}  \\
\beta &= 4\pi R^3 \frac{(\mu_r-1)(\eps_r+2) - \kappa^2}{(\eps_r+2)(\mu_r+2) - \kappa^2}  \\
\chi &= 12\pi R^3 \frac{ \kappa}{(\eps_r+2)(\mu_r+2) - \kappa^2} ~ .
\end{align*}
Here, the relative electric permittivity $\eps_r=\eps/\eps_0$ and magnetic permeability $\mu_r=\mu/\mu_0$ are defined with respect to vacuum, but the later result can be generalized to a spherical particle immersed in a non-dissipative medium with real parameters $\eps_m,\mu_m$ by redefining the relative permittivity and permeability with respect to the surrounding medium.

\bibliography{biblio_chiralmultipolar}

\end{document}